\documentclass{iopart}
\usepackage{a4}
\usepackage{graphicx}
\usepackage{subfigure}

\newcommand{\ds}{\displaystyle}
\newcommand{\scs}{\scriptscriptstyle}   

\def\o{\omega}
\def\d{\delta}
\def\g{\gamma}
\def\G{\Gamma}
\def\a{\alpha}
\def\b{\beta}
\def\l{\lambda}
\def\L{\Lambda}
\def\s{\sigma}

\def\th{\theta}
\def\dg{\dagger}
\def\la{\langle}
\def\ra{\rangle}

\def\pr{\scs \prime}
\begin{document}

\title[Interatomic potentials from EXAFS]
       {A quantum perturbative pair distribution for 
       determining interatomic potentials from EXAFS}

\author{F. Piazza$\dagger$}
\address{
   \dag INFM - UdR Firenze,  Via G. Sansone 1, 50019 
   Sesto F.no (FI), ITALY }
   

\begin{abstract}
In this paper we develop a technique for determining 
interatomic potentials in materials in the 
quantum regime from single--shell 
Extended X--ray Absorption Spectroscopy (EXAFS) spectra. 
We introduce 
a pair distribution function, based on ordinary
quantum time--independent perturbation theory.
In the proposed scheme, the model potential parameters enter the 
distribution through a fourth--order Taylor expansion of the 
potential, and are directly refined in the fit of
the model signal to the experimental spectrum.
We discuss in general the validity of our theoretical framework,
namely the quantum regime and perturbative treatment,
and work out a simple tool for monitoring the sensitivity
of our theory in determining lattice anharmonicities based on
the statistical $F$--test.
As an example, we apply our formalism to an EXAFS spectrum at the Ag K--edge
of AgI at $T=77$ K. We determine the Ag--I potential parameters
and find good agreement with previous studies.
\end{abstract}
\section{Introduction}
It is well known that EXAFS is a very sensitive and accurate 
technique for probing the structural and dynamical properties
of materials in the neighborhood of the photoabsorber 
atom~\cite{koning}. In particular, the damping of the EXAFS
signal induced by thermal broadening in the distribution of 
absorber--neighbour distances carrys quantitative information
on the corresponding interatomic potentials~\cite{teo}.

The effects of thermal disorder are usually accounted for
by introducing a temperature--dependent pair distribution function (PDF) 
$g(r,T)$, such that $g_s(r,T)\,dr$ is the normalised probability that 
the distance of an atom in the the $s$--th shell from the 
absorber lies in the interval  [$r$,$r+dr$] at the temperature $T$.
The corresponding single--shell EXAFS $\chi_s(k)$ is then obtained as
\begin{equation}
\label{chi}
\chi_s(k) = \int_0^\infty g_s(r,T) \chi_s(k,r) \, dr \quad .
\end{equation}
It is clear that the absorber--neighbour 
effective potential determine the shape of the PDF.
The simplest structural model one can introduce is represented
by a Gaussian distribution, i.e. harmonic interatomic potential. 
In this case, integration of eq.~(\ref{chi}) is straightforward 
and the result is the well--known Debye--Waller multiplicative damping
factor  $e^{-2 \s^2 k^2}$,  where 
$\s^2 = \langle [ r - \langle r \rangle ]^2 \rangle
= \int_0^\infty g(r,T) [ r - \langle r \rangle ]^2 dr$.

The higher--order terms in the potential expansion 
are usually accounted for via the cumulant method~\cite{bunker}.
The cumulants $C_n$ are defined as the coefficients that
enter the MacLaurin expansion of the function $\ln[F(2k)]$,
where $F(2k)$ is the Fourier transform of the 
effective distribution function 
$G(r,\lambda) = g(r)\,e^{-2r/\lambda} / r^2$,
$\lambda$ being the photoelectron mean free path.
The utility of this method is that the cumulants are related
to the moments of the effective distribution.
In particular, the first cumulant $C_1$ is the 
mean value of the interatomic distance, while 
$C_2$ is the variance of the effective distribution. The higher--order 
cumulants $C_3$ and $C_4$ are related to the {\em skewness} (asimmetry) 
and {\em kurtosis} (deviation from the Gaussian shape) of the distribution,
respectively. 
The cumulant--expansion technique has been applied succesfully
in the case of moderate anharmonicities both to bulk 
materials~\cite{cum_bulk} and to the study of 
surfaces~\cite{cum_surf1,cum_surf2}.

In such a framework, the following step towards a detailed understanding 
of disordered systems  is to establish the direct relationship between 
the interatomic potential and cumulants. In the case of harmonic
crystals, the second--order cumulants have been calculated
quantum--mechanically for Debye crystals~\cite{cum_Debye} and
simple molecules~\cite{cum_molecule}. Other 
attempts in this direction exist in the literature
which include
third--order anharmonicities for simple systems~\cite{Rabus,FrenkRehr}. 
 
A more straightforward alternative approach is to directly integrate the 
EXAFS by calculating explicitely the whole pair distribution function
corresponding to a certain model potential.
The simplest way a pair potential $V(r)$
may directly enter the distribution function is through the classical
configurational integral scheme
\begin{equation}
\label{grB}
g(r,T)=\frac{\ds e^{\ds -V(r)/k_{\scs B} T}}
            {\ds \int_0^\infty e^{\ds -V(r)/k_{\scs B} T} \, dr} \quad .
\end{equation}
where $k_{\scs B}$ is the Boltzmann constant.
Expression~(\ref{grB}) has been applied to the study of 
metals and ionic systems with different model
potentials, from Lennard--Jones~\cite{LJ_pot}, Morse~\cite{Morse_pot}
to generic three--parameter Taylor expansions~\cite{Tayl_pot}.
However, eq.~(\ref{grB}) is based on a classical treatment of
the atomic vibrations. In general, depending both on
the temperature and potential stiffness, 
the classical approximation may break down. In this case the 
full quantum treatment of lattice dynamics is in  order, and
one can proceed as follows. 

First the Schr\"{o}dinger equation of the absorber--neighbour pair
has to be solved, and the eigenvectors $\psi_n$  and 
eigenvalues $E_n$ computed. 
The quantum--mechanical analogue of expression~(\ref{grB}) can
then be written as
\begin{equation}
\label{grQ}
g(r,T)=\frac{\ds \sum_n |\psi_n(r)|^2 e^{-E_n/k_{\scs B} T}}
          {\ds \sum_n e^{-E_n/k_{\scs B} T}} \quad ,
\end{equation}
where $r=|{\bf x}_1 - {\bf x}_2|$ is the spatial coordinate 
describing the relative motion of the pair.
Eq.~(\ref{grQ}) has been first introduced and used in ref.~\cite{Bishop} 
to study the interatomic potential of the Cu--O(4) 
pair in the YBCO superconductor. 
The procedure is the following: one first 
introduces a model potential which is characterised by a set of parameters 
$\{ \l \}$. Then a numerical routine is set up, which solves the 
radial Schr\"{o}dinger equation for a particular choice of the set $\{ \l \}$,
builds the radial distribution function~(\ref{grQ})
and calculates the EXAFS signal $\chi(k,\{ \l \})$. This routine is then
incorporated in the fitting program that refines the free parameters
$\{ \l \}$ on a set of experimental data  through ordinary 
$\chi^2$ minimisation.

This procedure has the advantage that it allows an arbitrary 
analytical potential function to be used. This is the case of 
ref.~\cite{Bishop}, where a double--well potential is found.
However, it is rather cumbersome and of little practical
utility for routine fittings.
In particular, a more concise and handy 
way of calculating expression~(\ref{grQ}) in a closed form 
would be of great advantage.

In this paper we compute an analytical expression for 
the function~(\ref{grQ}), based on a simple Taylor expansion of
the potential function, and cast it in a simple
form, suitable for inclusion in a simple routine attached to the
fitting machinery.  We organise our paper as follows.
In section~\ref{sec2} we develop our quantum pair
distribution function (QPDF). In section~\ref{sec3} we discuss
on general grounds the validity of our theoretical framework. 
Furthermore, we develop a  statistical tool to assess the sensitivity of 
the QPDF to the parameters describing the anharmonicities
in the potential. 
Finally, in section~\ref{sec4} we test the QPDF by fitting an 
EXAFS spectrum of AgI at $T=77$ K. We end the paper by summarising
our results and drawing our conclusions in section~\ref{sec5}. 

\section{\label{sec2} The Quantum Pair Distribution Function}
Let us consider the pair formed by the photoabsorber and 
one of its neighbours from a given coordination shell. 
Let $\bf{x}_i$ and $m_i$ ($i=1,2$) denote the position vectors
and atomic masses of the two atoms, respectively. Let 
$V(r)$ be the corresponding interatomic potential.
We can write its Taylor expansion in the following fashion
\begin{equation}
\label{Taylpot}
V(r)=\frac{1}{2} k_2 (r-r_0)^2 + V_1(r) + {\cal O}(|r-r_0|^5)  \quad ,
\end{equation}
where 
\begin{equation}
\label{V1}
\fl
V_1(r) = \frac{1}{3} k_3 (r-r_0)^3 +
     \frac{1}{4} k_4 (r-r_0)^4 
     \ \ \ \ \ {\rm and}  \ \ \ \ \ 
     k_m = \frac{1}{(m-1) !} \left[ 
                         \frac{d^{\, m} V(r)}{d r^m} 
                         \right]_{r=r_0} \quad ,
\end{equation}
$r_0$ being the equilibrium interparticle distance, given by 
the condition $[ d V(r) / d r]_{r=r_0} = 0$.
We can follow the ordinary procedure to first separate the two--body  
Schr\"{o}dinger equation by introducing the relative and 
centre--of--mass coordinates, and then decouple the angular
and radial degrees of freedom in the Schr\"odinger
equation for the radial motion. The pair wavefunction then reads
\begin{equation}
\label{wf}
\Psi({\bf x}_1,{\bf x}_2) = \psi_{\scs G}({\bf X}) 
                          \left[
			  \frac{u(r)}{r}
			  \right]
			  Y^m_l(\th,\phi) \quad  ,
\end{equation}
where ${\bf X} = (m_1 {\bf x}_1 + m_2 {\bf x}_2)/(m_1+m_2)$, 
$r = |{\bf x}_1 - {\bf x}_2|$ and $Y^m_l(\th,\phi)$ are the
spherical harmonic functions~\cite{winter}.
We require the wavefunction of the pair to have spherical symmetry, 
since we do not want the PDF to depend on 
the orientation of the absorber--neighbour bond in the crystal
\footnote{this also means that our formalism in its
present form only applies to the study of K--edges.}. 
We therefore set $l=0$.
The radial equation then reduces to the one--dimensional problem
\begin{equation}
\label{rad_expl}
- \frac{\hbar^2}{2 \mu} \frac{d^{\, 2} u}{d r^2} +
                \left[
                \frac{1}{2} k_2 (r-r_0)^2 + V_1(r) 
                \right] u = E u \quad ,
\end{equation}  
where $\mu$ is the reduced mass of the pair and 
we require $u(r)|_{r=0} = 0$.
In the spirit of ordinary time--independent perturbation theory, 
we consider the harmonic Hamiltonian as the unperturbed problem 
and the potential $V_1$ as the perturbation.

It is convenient to adopt the formalism 
of second quantisation.
The unperturbed problem is defined by the eigenvectors $| n \ra$
and the corresponding eigenvalues 
$E^{(0)}_n = \hbar \o [ n + 1/2] \ (n=0,1,2, \dots)$, 
where $\o = \sqrt{k_2 / \mu}$.
Recalling the well known commutation relations between creation and 
annihilation operators
$\hat{a}^\dg$ and  $\hat{a}$, respectively,
it is straightforward to write down the expression of the perturbation
potential~(\ref{V1}) in the $n$--representation. We get
\begin{eqnarray}
\label{Vpertn}
\fl
{\cal V} =  \hbar \o \, \L_3 [\hat{a}^{\dg 3} + \hat{a}^3 + 3 n \hat{a}^\dg + 
                                  3(n+1) \hat{a}] \, + \\ \nonumber
            \hbar \o \, \L_4 [\hat{a}^{\dg 4} + \hat{a}^4 +
                      2(2n-1) \hat{a}^{\dg 2} + 2(2n+3) \hat{a}^2 
                      +3 (n+1)^2 + 3 n^2] \quad ,
\end{eqnarray}
where
\begin{equation}
\label{Lambda}
\Lambda_m = \frac{1}{2^{m/2} m} \left( 
                                \frac{k_m x_0^m}{\hbar \o} 
                                \right)  \ \ (m=3,4) \ \ \ \ \ \ 
				{\rm and} \ \ \ \ \ \
				x_0=\sqrt{\hbar / \mu \o} \quad .
\end{equation}
Recalling the definition of creation 
and annihilation operators
we can easily evaluate the matrix elements $\la k | {\cal V}| n \ra$. 
Let us define the four--dimensional quantity $\b$ as
\begin{equation}
\label{beta}
\b(\s) = \L_3 [\d_{\s,1} + \d_{\s,3}] + 
         \L_4 [\d_{\s,2} + \d_{\s,4}] \ \ (\s=1,2,3,4) \quad ,
\end{equation}
where $\d_{i,j}$ is the Kronecker integer delta function.
We get
\begin{equation}
\label{Vkn}
\fl
\la k | {\cal V} | n \ra = \la n | {\cal V} | k \ra =
\sum_{\s=1}^4 \b(\s) \left [
                     \g_n^{+}(\s) \d_{k,n+\s} + 
                     \g_n^{-}(\s) \d_{k,n-\s}
                     \right] \quad ,
\end{equation}
where we have explicitly used the hermiticity of ${\cal V}$, and
defined
\begin{equation}
\label{gamma}
\fl
\g_n^{+}(\s) = \sqrt{\frac{(n+\s)!}{n!}} \a^{+}(n,\s) \ \ \ \ \ \ \ 
\g_n^{-}(\s) =
                        \left\{
                        \begin{array}{ll}
\sqrt{\frac{\ds n!}{\ds (n-\s)!}} \a^{-}(n,\s) & n \geq \s \\
0                                      & {\rm otherwise} .
                         \end{array}
                         \right.
\end{equation}
The coefficient $\a^\pm(n,\s)$ are reported in table~\ref{tab_alph}.
We note that the coefficients 
$\g_n$ and $\a^\pm(n,\s)$ satisfy to the 
following relations
\begin{equation}
\label{cond_gam}
\begin{array}{llll}
\g_{n-\s}^{+}(\s) &= \g^{-}_n & \qquad
\a^{+}(n-\s,\s)   &= \a^{-}(n,\s) \\
\g_{n+\s}^{-}(\s) &= \g^{+}_n & \qquad
\a^{-}(n+\s,\s)   &= \a^{+}(n,\s) \quad .
\end{array}
\end{equation}
Using eq.~(\ref{Vkn}),~(\ref{gamma}) and~(\ref{cond_gam}), 
we can work out the second--order 
corrections to the energy levels $\Delta E_n$ and
the normalised perturbed wavefunctions $| n \ra + | n^{\pr} \ra$
in the usual way
(see e.g. ref.~\cite{cohentann}).
After a somewhat lengthy calculation, we get
\begin{eqnarray}
\label{DE_sum}
\fl
\Delta E_n/\hbar \o = 3 \L_4 (2n^2 +2n+1) 
             + \sum_{\s=1}^4 \frac{\b(\s)^2}{\s}
                    \left[
                      (\g_n^{-}(\s))^2  - (\g_n^{+}(\s))^2
                    \right] \quad .
\end{eqnarray}
\begin{eqnarray}
\label{Psi12f}
\fl
| n^{\pr} \ra  = \sum_{\s=1}^4 \frac{\b(\s)}{\s}
                \left[
                 \g_n^{-}(\s) | n - \s \ra - 
                  \g_n^{+}(\s) | n + \s \ra
                 \right]  + \nonumber \\
	       \fl  \sum_{\s, \s^{\pr}=1}^4
                \frac{\b(\s) \b(\s^{\pr})}{\s (\s + \s^{\pr})}
                \left[
                  \g_n^{+}(\s) \g_{n+\s}^{+}(\s^{\pr})
                     | n + ( \s + \s^{\pr}) \ra
               +  \g_n^{-}(\s) \g_{n-\s}^{-}(\s^{\pr})
                     | n - ( \s + \s^{\pr}) \ra
                \right] - \nonumber \\
              \fl  \sum_{\s \neq \s^{\pr}=1}^4
                \frac{\b(\s) \b(\s^{\pr})}{\s (\s^{\pr} -\s)}
                \left[
                  \g_n^{+}(\s) \g_{n+\s}^{-}(\s^{\pr})
                     | n - ( \s^{\pr} - \s ) \ra 
               +  \g_n^{-}(\s) \g_{n-\s}^{+}(\s^{\pr})
                     | n + ( \s^{\pr} - \s ) \ra
                \right] - \nonumber \\
               \fl  \frac{\Delta E_n^{(1)}}{\hbar \o} \sum_{\s=1}^4
                 \frac{\b(\s)}{\s^2}
                 \left[
                  \g_n^{+}(\s) | n+\s \ra + \g_n^{-}(\s) | n-\s \ra
                 \right] - \nonumber \\ 
                 \frac{1}{2} | n \ra \sum_{\s=1}^4
                 \frac{\b^2(\s)}{\s^2}
                 \left[
                  (\g_n^{+}(\s))^2  + (\g_n^{-}(\s))^2 
                 \right] \quad ,	 
\end{eqnarray}
where $\Delta E_n^{(1)} = 3 \hbar \o \L_4 (2n^2 +2n+1) $ are the first--order 
corrections to the energy levels.

We are now able to write down explicitly  the expression for the QPDF 
from eq.~(\ref{grQ}). We have
\begin{equation}
\label{grQ1}
g(r,T)=\frac{\ds \sum_{n=0}^{n_{\scs M}} 
      [u_n(r) + u^{\pr}_n(r)]^2 
      e^{\ds -[E^{(0)}_n + \Delta E_n] / 
      k_{\scs B} T}}
           {\ds \sum_{n=0}^{n_{\scs M}} 
      e^{\ds -[E^{(0)}_n + \Delta E_n] / 
      k_{\scs B} T}} \quad ,
\end{equation}
where $u_n(r) = \la r | n \ra $ are the unperturbed eigenfunctions, 
i.e. the eigenvectors of the one--dimensional harmonic oscillator,
and $u^{\pr}_n(r) = \la r | n^{\pr} \ra $ are the corrections~(\ref{Psi12f}).
We have explicitly indicated the truncation of the summations as
the integer $n_{\scs M}$. In the computations one has to fix
$n_{\scs M}$ by requiring that the corresponding normalised Boltzmann factor 
$z_{n_{\scs M}}/Z$ ($Z$ being the partition function)
is negligible up to some specified tolerance $tol=10^{-M}$, i.e.
\begin{equation}
\label{z_tol}
z_{n_{\scs M}} = e^{\ds -[E^{(0)}_{n_{\scs M}} + 
              \Delta E_{n_{\scs M}}] / 
              k_{\scs B} T} \le tol \quad .
\end{equation}
This condition fixes the number of levels which are included
in the perturbative series~(\ref{grQ1}).
Of course, one also has to check that the energy
of the highest level included is small compared to
some estimate of the the potential well depth $V_o$.
If we express energies
in eV and temperature in Kelvin, from eq.~(\ref{z_tol}) we get
the condition
\begin{equation}
\label{tol_well}
[M \log 10 ] \, T  \times  10^{-4} < V_o \quad .
\end{equation}
It follows that, for potential well depths of the order of 1 eV,
the condition~(\ref{tol_well}) is fulfilled for $M=4-5$ for
temperatures up to $\approx 100$ K. 
\section{\label{sec3} Validity and sensitivity of the QPDF}
The above described procedure to build the QPDF relies on two basic 
hypotheses: $(i)$ the classical approximation
of lattice vibrations must be inadequate so that the quantum treatment 
of the two--body problem holds, and $(ii)$ the deviations
of the absorber--neighbour potential from the harmonic approximations
must be satisfactorily described by a perturbative treatment.
As to the validity of condition $(i)$, the ratio 
${\cal R}_{\scs Q} = \hbar \o / k_{\scs B} T$ 
($\o = \sqrt{k_2 / \mu}$) provides a good 
qualitative indicator: if ${\cal R}_{\scs Q}$ is of order one, 
the quantum energy scale is comparable with the thermal one, 
and the system requires the full quantum description.
Regarding condition $(ii)$, we introduce the following 
parameter
\begin{equation}
\label{Hpert}
\fl
{\cal R}_{\scs E} \stackrel{def}{=} 
                   \frac{2 V_1 (\sqrt{\la (r-r_0)^2 \ra_T})}
                          {k_2 \la (r-r_0)^2 \ra_T } 
                  = \frac{1}{k_2} \left[
                     \frac{2}{3} k_3 \sqrt{\la (r-r_0)^2 \ra_T} +
                     \frac{1}{2} k_4 \la (r-r_0)^2 \ra_T
                                     \right] \quad ,
\end{equation} 
where $\la \dots \ra_T$ is the configurational average 
performed with the QPDF.
The indicator ${\cal R}_{\scs E}$ gives
a measure of the relative strength of the perturbed and 
unperturbed energies, and can always be 
computed {\em a posteriori} in order to assess the validity
of the perturbative approximation.
On the other hand, we also have to be concerned with the {\em sensitivity} of 
the QPDF to the parameters of the model potential. i.e. the minimum detectable 
anharmonicity within the present model at fixed temperature and
potential stiffness ($k_2$). We shall here introduce 
a simple procedure for assessing the QPDF sensitivity, based
on the statistical $F$--test.

Let us suppose that we are fitting $N$ experimental data points 
$(k_i,\chi^{\scs exp}_i)$, $i = 1, \dots, N$, 
to a model that has $p$ adjustable parameters $\l_j$, $j = 1, \dots, p$. 
The model predicts a functional
relationship between the measured independent and dependent variables
\[
\chi(k) = \chi^{\scs the}(k; \l_1, \dots ,\l_p) \quad .
\]
In the spirit of the maximum--likelihood 
method, we want to minimise a fit index (or residual function) of the kind
\begin{equation}
\label{resid}
{\cal F} = \sum_{i=1}^N [\chi^{\scs exp}_i - 
                      \chi^{\scs the}(k_i;\l_1, \dots ,\l_p) ]^2 \, w_i
\end{equation} 
where the $w_i$'s are some weight functions. In general, if the standard
deviations $\s_i$ of the experimental data are known independently, 
the correct choice would be $w_i \propto 1/\s_i^2$.
However, depending on the particular algorithm used to perform
the fit, some other weighting functions may be preferred.

As a consequence of introducing the third-- and fourth--order nonlinearities
in the model potential,  two more floating parameters are available to 
fit the EXAFS spectrum, namely $k_3$ and $k_4$. In general, this will cause 
{\em per se} a reduction of the best--fit index minimum with respect to the
harmonic model.  Such a situation typically arises in EXAFS data analysis 
when it is to be decided whether a spectrum needs the introduction 
of an additional shell to be fitted (some physical information still 
present in the data) or, more generally, whether the improvement achievable 
by incrementing  the number of free parameters is statistically 
meaningless~\cite{FtestEXAFS1,FtestEXAFS2}. 
In our case, we are interested in assessing the sensitivity of the
QPDF in capturing real physical information regarding
the higher--order terms in the Taylor expansion of the potential.

 A general theorem in statistics states that the minimum of the
residual function ${\cal F}_{\scs min}$ is distributed as a $\chi^2$
distribution with $\nu_1 = N - (p+1)$ degrees of freedom~\cite{Roe}
\footnote{This result strictly holds when $(i)$ the measurement errors
are normally distributed, and either $(ii)$ the model is linear in its parameters
or $(iii)$ the sample size is large enough that the uncertainties in the fitted
parameters do not extend outside a region in which the model could be 
replaced by a suitable linearised model. Fits in EXAFS are usually
at the limit of validity of condition $(iii)$. However, all the statistical
analysis that stems from this basic theorem is routinely applied
in EXAFS data analysis (see e.g.~\cite{fit_nonlin}).}.
Let us assume we want to compare an harmonic model potential 
(fit with $p$ parameters) to a potential
obtained by adding some anharmonicity (fit with $p+p^{\pr}$ parameters,
$p^{\pr}=1,2$). We want to assess whether
the latter model {\em significantly} improves the fit (given the
automatic improvement following the introduction of any additional
free parameter). 
Let us consider the ratio of the normalised minima of the two corresponding
fit indeces
\begin{equation}
\label{ratio}
F = \frac{{\cal F}_{\scs min}(p)/\nu_1}{{\cal F}_{\scs min}(p+p^{\pr})/\nu_2} \quad ,
\end{equation}
where $\nu_2 = [N-(p+p^{\pr}+1)]$. It can be shown that, as a consequence 
of the above theorem,  the ratio~(\ref{ratio}) 
follows an $F_{\scs \nu_1,\nu_2}$ distribution~\cite{Roe}, 
whose density function is
\begin{equation}
\label{densF}
{\cal D}_{\scs \nu_1,\nu_2}(F) dF = \frac{\ds \G[(\nu_1+\nu_2)/2] 
                                        [(\nu_1/\nu_2) F]^{\nu_1/2-1}}
                                        {\ds \G(\nu_1/2) \G(\nu_2/2) 
					[(\nu_1/\nu_2) F +1]^{(\nu_1+\nu_2)/2}} \, dF
					\quad .
\end{equation}
In particular, if both the harmonic and anharmonic models were appropriate 
to explaining all of the signal, one would expect the function
\begin{equation}
\label{ratio12}
f = \frac{\nu_2}{p^{\pr}} \left( \frac{{\cal F}_{\scs min}(p)}
                          {{\cal F}_{\scs min}(p + p^{\pr})} -1 \right) 
\end{equation}
to follow an $F_{\scs p^{\pr},\nu_2}$ distribution. 
The F--test is then conducted as  follows. (1) Based on some estimate 
of the experimental standard deviation, 
one generates an {\em artificial} experimental data set from the anharmonic 
model,  e.g. by adding some Gaussian noise. (2) The two residuals 
${\cal F}_{\scs min}(p)$ and  ${\cal F}_{\scs min}(p + p^{\pr})$  
are calculated and the value of 
$f$ obtained.  (3) One can now fix the preferred 
confidence level $c$ and compare $f$ with $F^c_{\scs p^{\pr},\nu_2}$,
given by the following relation
\begin{equation}
\label{c_ki}
c = 1 - \int_0^{F^c_{\scs p^{\pr},\nu_2}}
       {\cal D}_{\scs p^{\pr},\nu_2}(F) \, dF \quad .
\end{equation}
Here $c$ is the percentage 
probability of obtaining a reduction 
${\cal F}_{\scs min}(p) - {\cal F}_{\scs min}(p + p^{\pr})$
as large as actually observed, when the added anharmonicity is not
physically meaningful. 
We shall call such confidence level the {\em rejection probability},
which expresses the probability that the harmonic and anharmonic 
model are equivalent -- that is, if $c=1$ the two models are by all
means statistically indistinguishable, while if $c=0$ the
probability that the harmonic model could explain the data instead of the 
anharmonic one vanishes identically.

\section{\label{sec4} The Ag--I potential in Silver Iodide}
Silver iodide (AgI) is known for being a 
highly--anharmonic material~\cite{Yoshi1,Dalba1,Dalba2}.
Hence, it appears a good candidate for providing a bench--mark
to test the above described data analysis framework.
As an example, we analyse here an Ag K--edge EXAFS spectrum collected 
at $T=77$ K. The details of the experiment and of the extraction of
the EXAFS signal $\chi^{exp}(k)$ from the raw absorption data 
are reported  elsewhere~\cite{paolo_AgI}.

The classical expression~(\ref{grB})
of the PDF has been recently used in an EXAFS study
of AgI at the I K--edge 
to measure the first  three coefficients of the Taylor expansion
of the Ag--I potential at $T=300$ K and $T=600$ K~\cite{Yoshi2}.
Following ref~\cite{Yoshi2}, we get
${\cal R}_{\scs Q}(T=77) \approx 0.5$. We conclude that at $T=77$ K
the classical expression of the PDF is non longer valid and one
has to work in the quantum regime. 
In order to calculate the model EXAFS signal  we re--write 
eq.~(\ref{chi}) 
by using the standard formula of single--shell single--scattering 
EXAFS in the  following fashion
\begin{equation}
\label{standard}
\fl
\chi(k,r)= S_o^{2} N_{\rm I} \int_{-\infty}^{+ \infty} g(u,T)
            \frac{e^{-2[r_0+u]/\lambda(k)}}{k[r_0+u]^2}
            \,Im \left\{ 
	    f_{\rm I}(\pi,k)\,e^{2i\delta_{\rm Ag}}\,e^{2ik[r_0+u]} 
	          \right\} \, du \quad ,
\end{equation} 
where $g(u,T)$ is the QPDF as calculated from eq.~(\ref{grQ1})
and we have made the substitution $u=r-r_0$.
Here $N_{\rm I}$ is the coordination number of the I ions,
which we fixed at the crystallographic value  $N_{\rm I}=4$,
$\delta_{\rm Ag}$ is the central atom phase shift and 
$f_{\rm I}(\pi,k) = |f_{\rm I}(\pi,k)| \exp (i \phi_{\rm I})$
is the complex backscattering amplitude of the I ions.
The constant $S_o^2$ is the usual reduction factor which accounts for the
inelastic losses, which we fixed at the value $S_o^2=0.73$~\cite{paolo_AgI}.
The photoelectron mean free path is here assumed to depend on
the wavevector $k$ as $\lambda = k/\eta$~\cite{teo}, 
where $\eta$ is a constant.
The backscattering amplitude $f_{\rm I}(\pi,k)$ and total phase
shift $\Delta \varphi = 2 \delta_{\rm Ag} + \phi_{\rm I}$ 
have been taken from the tables in ref.~\cite{teo}
(reproduced from calculations based on the Herman--Skillman
wavefunctions~\cite{teo_tab}). 

Following the arguments developed in section~\ref{sec3},
we carry out two separate non--linear least--square fittings, by comparing 
the experimental EXAFS $\chi^{exp}(k)$ with the model 
signal calculated by eq.~(\ref{standard}) from the two separate 
sets of floating parameters 
$\{ \l \} = \{ r_0, \Delta E_0, \eta, k_2 \}$ and 
$\{ \l^{\pr} \} = \{ r_0, \Delta E_0, \eta, k_2, k_3, k_4 \}$.
The floating parameter $\Delta E_0$ must be included in the fit 
as usual in order to compensate for the uncertainty 
associated with the {\em true} value of the threshold energy, 
i.e. the minimum energy required to free the photoelectron.

\begin{figure}
\centering
\subfigure[]{\includegraphics[width=6.5 truecm]{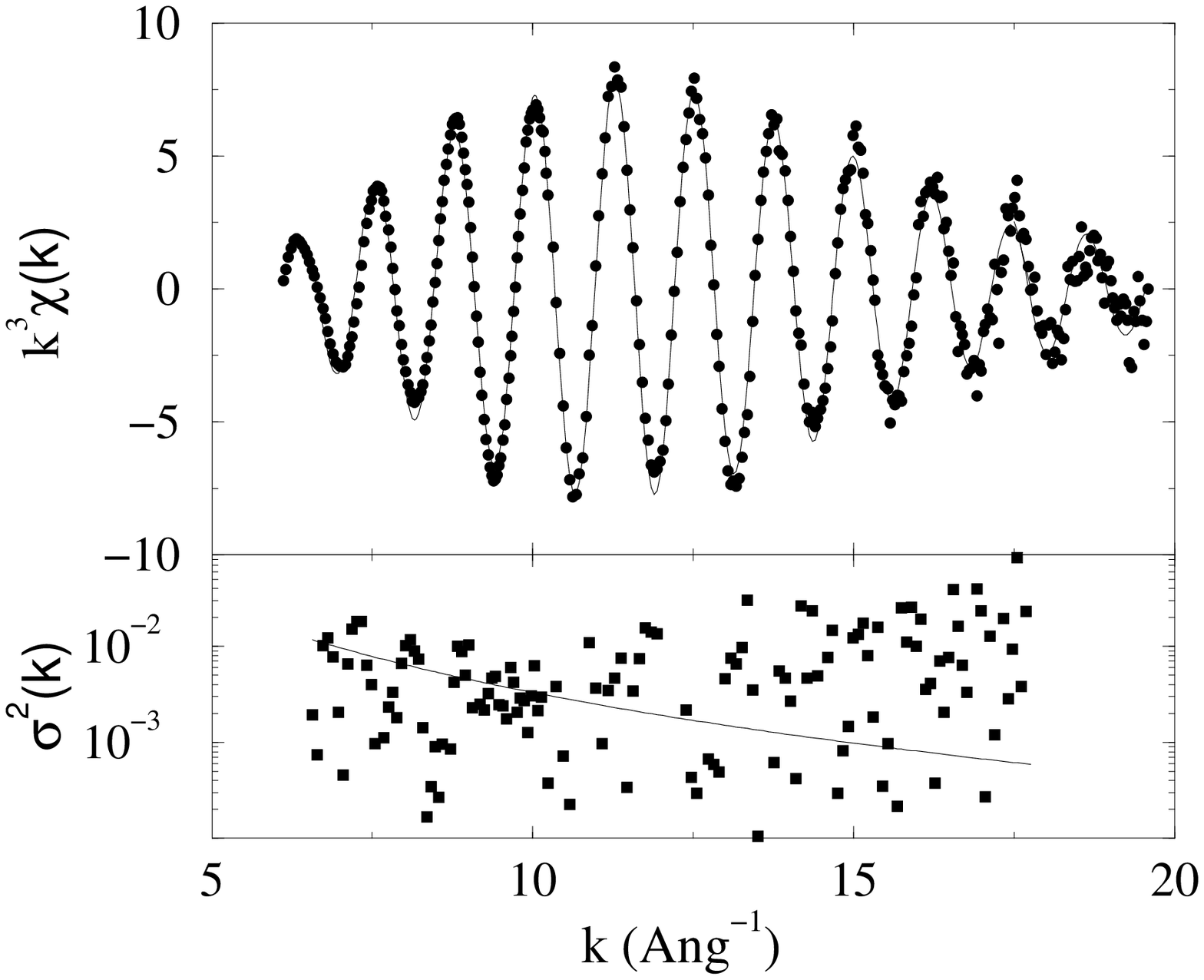}}
\subfigure[]{\includegraphics[width=6 truecm, height=5.4 truecm]
            {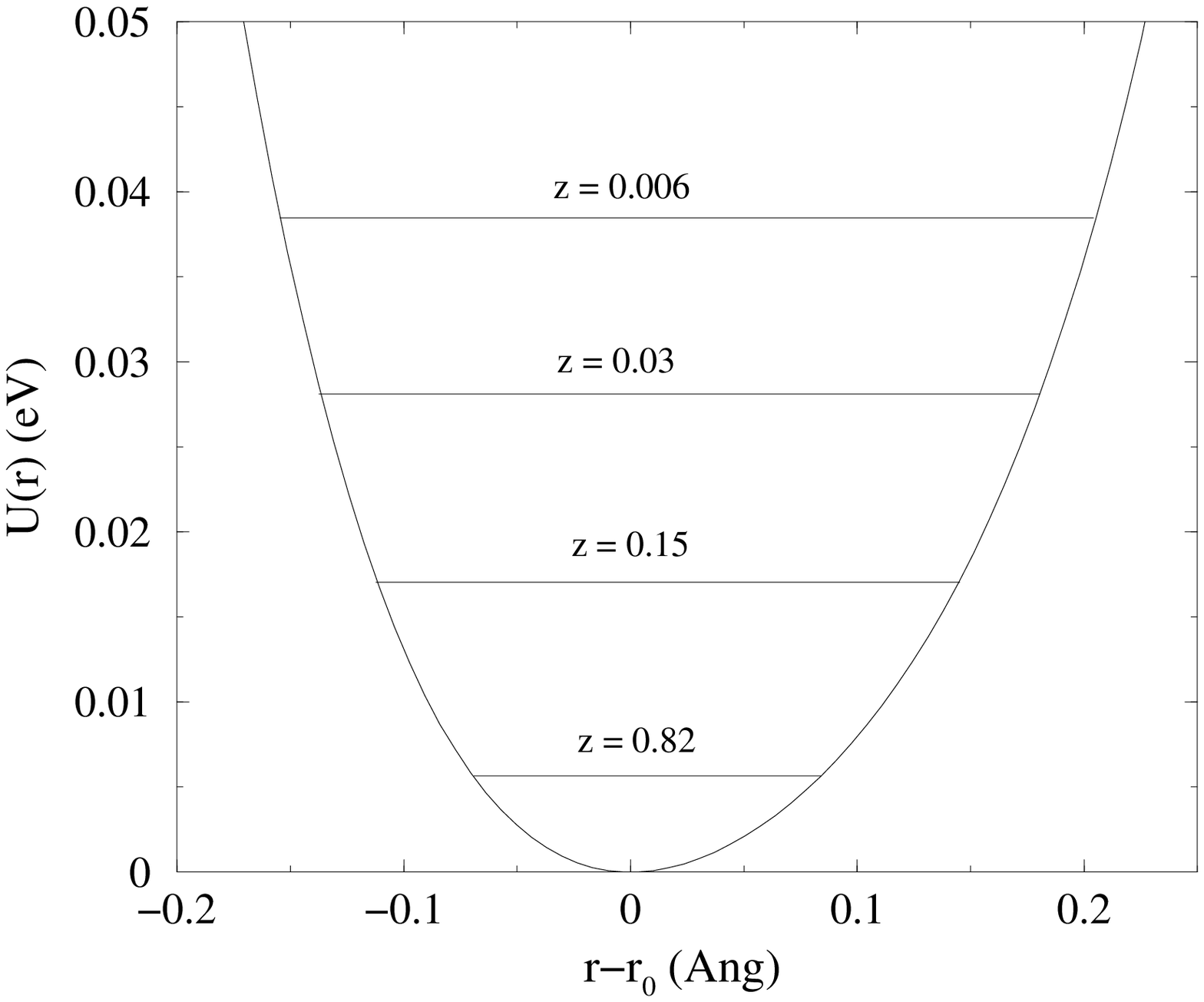}}
\caption{\label{fig_AgI_fit}  (a) Upper frame: Ag K--edge EXAFS signal
at $T=77$ K (symbols) and fit obtained with eq.~(\ref{standard})
and parameter set $\{ \l^{\pr} \}$.
Lower frame:
squared residuals (symbols) and fit with the law
$\s^2(k) = \s^2_0 \, [k_0/k]^3$ (solid line),
with $\s^2_0 = 0.0016$ and $k_0 = 12.7$ \AA$^{-1}$.
(b). Effective Ag--I potential.
Also shown are the harmonic levels used to calculate the QPDF, and the 
corresponding  normalised Boltzmann factors $z_n/Z$.}
\end{figure}

The best--fit values of the free parameters are reported in 
table~\ref{tab_AgI_mia} for both the harmonic and anharmonic models,
alongside with the corresponding fit index minimum ${\cal F}_{min}$.
The quality of the fit with the anharmonic model is shown in 
fig.~\ref{fig_AgI_fit} (a), while the corresponding effective Ag--I
potential is drawn in fig.~\ref{fig_AgI_fit} (b).
In table~\ref{tab_AgI} we report for comparison the values
of the potential parameters as measured in ref.~\cite{Yoshi2}.
We see that the overall agreement is good, although the 
value of $k_4$ reported in ref.~\cite{Yoshi2} does not seem 
to be reliable.  Moreover, if we use eq.~(\ref{Hpert}) to
estimate the strength of the perturbation energy corresponding 
to the best--fit values of $k_3$ and $k_4$, we get
$ {\cal R}_{\scs E} \approx 0.12$. We conclude that our results
of the analysis with the QPDF are consistent with the 
perturbative hypothesis. It is instructive to observe that
the same analysis performed using the classical expression
of the PDF eq.~(\ref{grB}) with the parameter set $\{ \l^{\pr} \}$
always yields a vanishing $k_2$ at the
minimum ${\cal F}_{min}$. Correspondingly, 
the fourth--order constant  $k_4$ is raised to
unphysically high values, independently of the initial
guess of the set $\{ \l^{\pr} \}$. 
This scenario clearly confirms 
the inadequacy of the classical treatment of the pair dynamics
in the present case.  

\begin{figure}[t!]
\centering
\includegraphics[width=8.5 truecm]{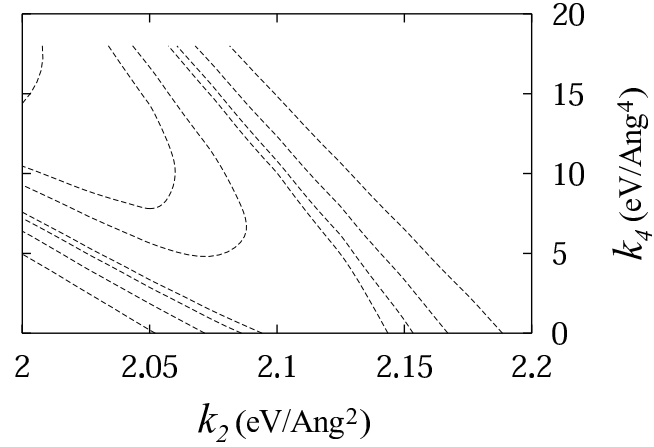}
\caption{\label{fig_chi2_AgIk40} Contour levels of the 
function $\Delta {\cal F} = {\cal F}(k_2,k_4) - {\cal F}_{min}$ for the 
minimum corresponding to the harmonic model. The contours correspond,
from left to right, to $\Delta {\cal F}$=$-[\chi^2]^{0.01}_{\nu=2}$,
$-[\chi^2]^{0.1}_{\nu=2}$,$-[\chi^2]^{0.3}_{\nu=2}$,
$[\chi^2]^{0.5}_{\nu=2}$,$[\chi^2]^{0.3}_{\nu=2}$, 
$[\chi^2]^{0.1}_{\nu=2}$ and $[\chi^2]^{0.01}_{\nu=2}$.}
\end{figure}

\begin{figure}
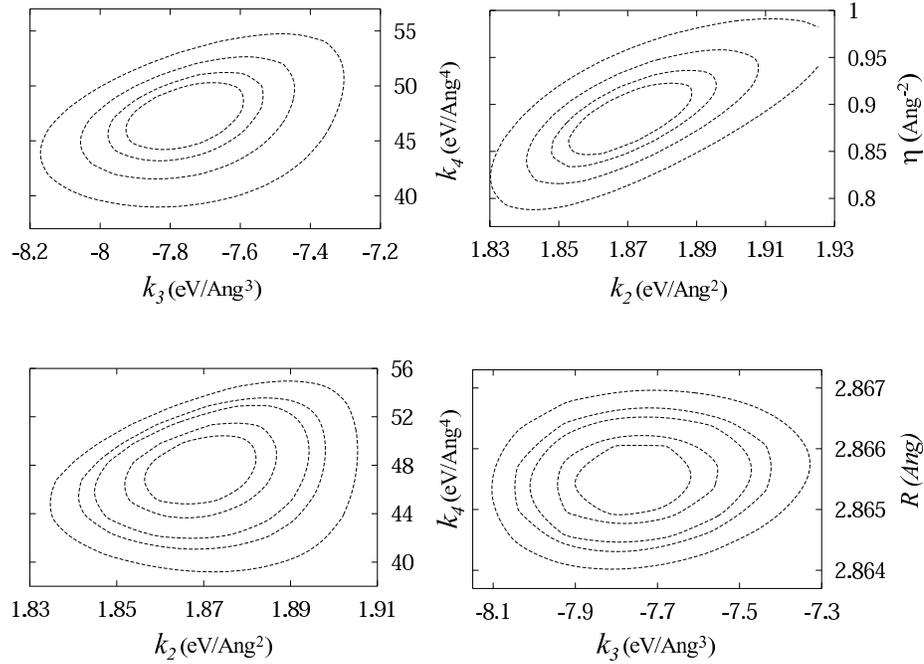

\centering
\subfigure{\includegraphics[width=6 truecm]{chi2_AgI77_k3k4.eps.new}}
\subfigure{\includegraphics[width=6 truecm]{chi2_AgI77_k2eta.eps.new}}
\subfigure{\includegraphics[width=6 truecm]{chi2_AgI_k2k4_77.eps.new}}
\subfigure{\includegraphics[width=6 truecm]{chi2_AgI_k3R_77.eps.new}}
\caption{\label{fig_AgI_cpl4}  
Contour levels of the function 
$\Delta {\cal F} = {\cal F}(\l_1,\l_2) - {\cal F}_{min}$ for the 
minimum corresponding to the anharmonc model. Four combinations
of floating parameters pairs $\{ \l_1, \l_2 \}$ are shown.
Contours correspond to confidence levels
$50 \%, 70 \%, 90 \%, (95 \%)$ and 99 \%.}
\end{figure}

We turn now to assessing the validity of our treatment on statistical
grounds.  The  value of $f$  corresponding to the  reduction of ${\cal F}$ 
following the introduction 
of $k_3$ and $k_4$ can be calculated by eq.~(\ref{ratio12}).
Substituting $N=150$ and $p^{\pr}=2$, we get $f \approx 16.6$.
By substituting in turn $F^c_{\scs p^{\pr},\nu_2}=f$
in eq.~(\ref{c_ki}) we get the corresponding rejection probability 
$c \approx 3 \times 10^{-7}$. We are then allowed to conclude 
that the anharmonic model is here capturing a real physical 
feature of the Ag--I pair dynamics.
It is instructive to demonstrate this conclusion in a more pictorial fashion.
It is a simple corollary to the first theorem mentioned in 
section~\ref{sec3} that, if only $\nu$ parameters are varied 
while keeping the other $p - \nu$  fixed at their best--fit value,
the function $\Delta {\cal F} = {\cal F}(\lambda_1,\lambda_2,\dots,\lambda_\nu) - 
{\cal F}_{min}$  follows a $\chi^2$ distribution  with $\nu$ degrees of freedom.
When $\nu=2$ this result provides the errors on selected 
parameter pairs $(\lambda_1,\lambda_2)$ in the form of 
{\em confidence ellipses} through the simple condition 
$\Delta {\cal F}(\lambda_1,\lambda_2) = [\chi^2]^c_{\nu=2}$.
Here $[\chi^2]^c_{\nu=2}$ is the value of the $\chi^2$ variable
corresponding to the required confidence level (i.e. rejection 
probability) $c$ for $\nu=2$.
In fig.~\ref{fig_chi2_AgIk40} we show the contour levels of the function 
$\Delta {\cal F} = {\cal F}(k_2,k_4) - {\cal F}_{min}$
computed for the minimum obtained within the harmonic model.
The presence of a significant region (corresponding to
the 99 \% confidence level $[\chi^2]^{0.01}_{\nu=2}=9.21$) 
of negative values away from the computed minimum
explains the dramatic improvement of the fit upon introducing
the nonlinearities in the potential. The same confidence analysis
for the anharmonic model performed
in four different parameter subspaces $\{ \l_1,\l_2 \}$
is reported in fig.~\ref{fig_AgI_cpl4},
showing the quality of the best--fit minimum.

\begin{figure}[t!]
\centering
\subfigure[]{\includegraphics[width= 6 truecm]{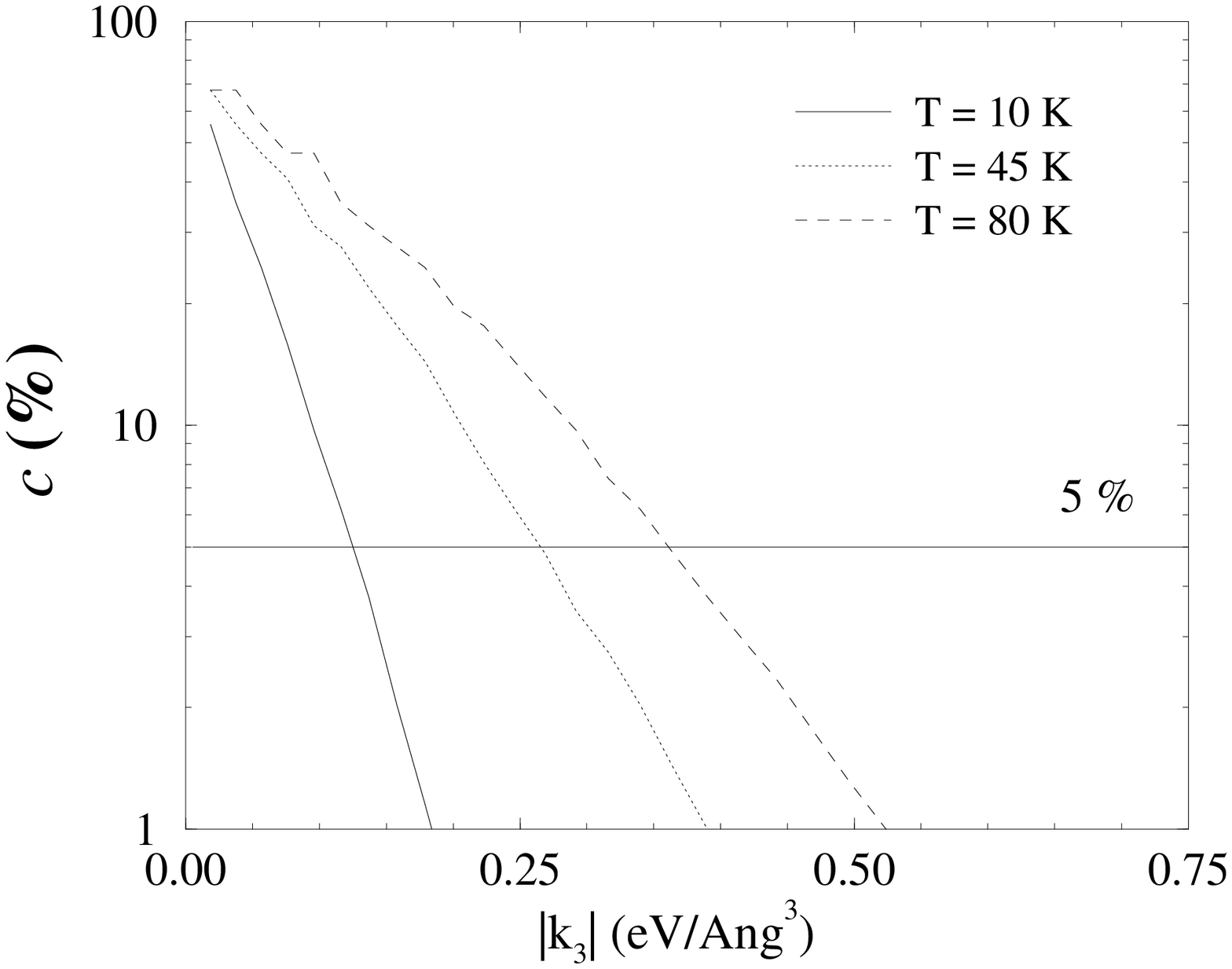}}
\subfigure[]{\includegraphics[width= 6 truecm]{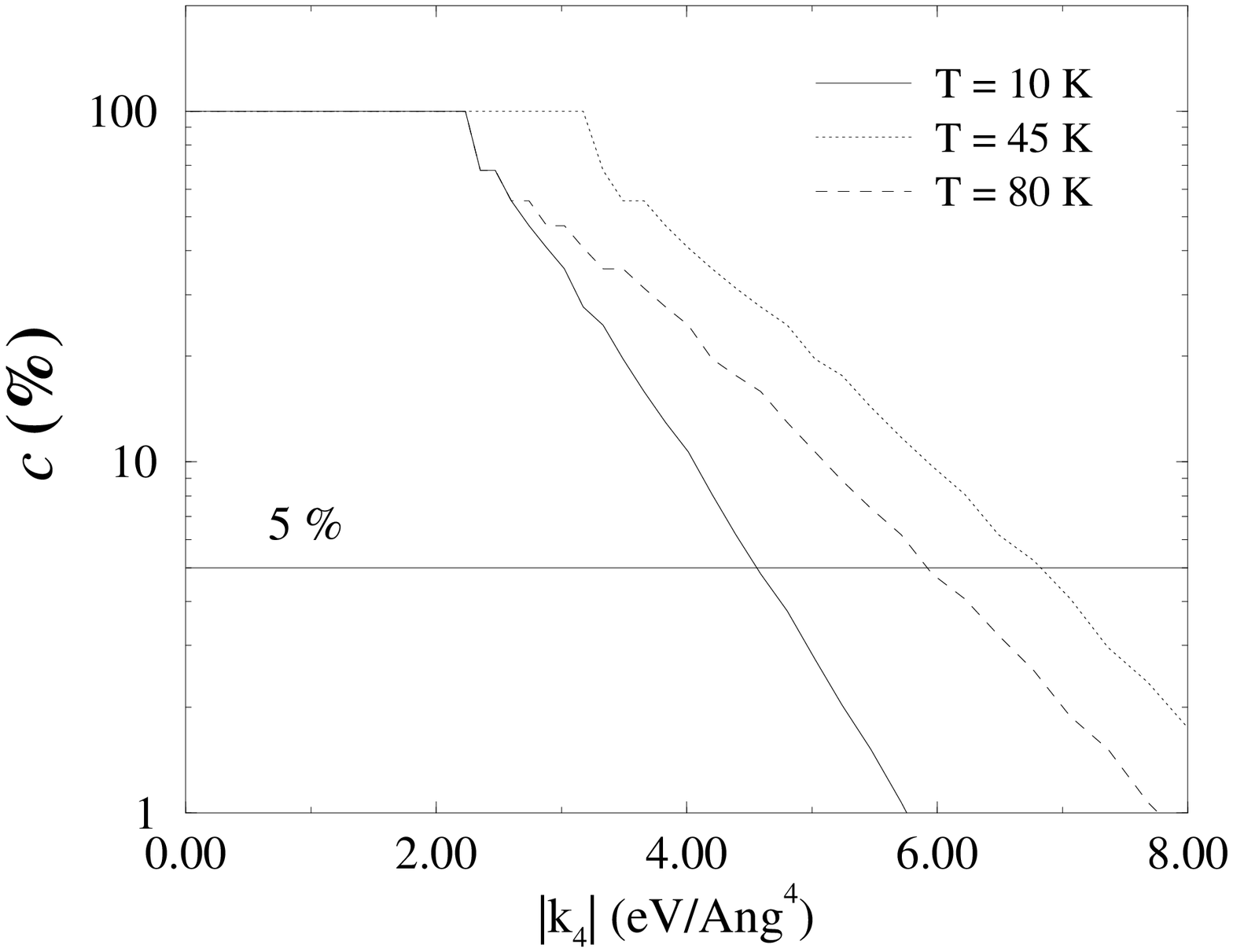}}
\caption{\label{prejX_k3k4} Rejection probability as 
a function of the potential anharmonicity calculated by eq.~(\ref{c_ki}) 
at different temperatures for fixed harmonic constant.
Parameters are $N=150$, $\mu=9.74 \times 10^{-26}$ Kg 
and $k_2=1.86$ eV/\AA$^2$. (a) cubic nonlinearity (b) quartic nonlinearity}
\end{figure}

We end this section by showing how one can conduct the $F$--test
described in section~\ref{sec3} to examine the QPDF sensitivity
in the present case.
Let us fix the temperature and the harmonic constant $k_2$ at its best--fit
value.
We can then use eqs.~(\ref{standard}),(\ref{resid}) and~(\ref{ratio12}) 
with $p^{\pr}=1$ to calculate the value of $f$ for any 
choice of $k_3$ and $k_4$, where in place of the experimental 
spectrum we use an artificial data set constructed as
described in section~\ref{sec3}. 
In particular, we just add to the model signal a Gaussian
noise with standard deviation
$\s(k,T) = \s_0 \, [k_0/k]^{3/2} \sqrt{T/T_0}$,
with $\s^2_0 = 0.0016$, $k_0 = 12.7$ \AA$^{-1}$ and 
$T_0=77$ K (see fig.~\ref{fig_AgI_fit} (a)).  Finally, we calculate the 
corresponding rejection probability  by means of 
eq.~(\ref{c_ki}) with $F^c_{\scs p^{\pr},\nu_2}=f$.
For example, this procedure can be used to construct the functions
$c(k_3,k_4=0,T)$ and $c(k_3=0,k_4,T)$
(one--parameter sensitivity curves). Alternatively, the same procedure 
with $p^{\pr}=2$ can be used to look at the contour sections 
of the function $c(k_3,k_4,T)$ (two--parameter sensitivity curves).  
In fig.~\ref{prejX_k3k4} we show an example of one--parameter
sensitivity curves calculated at three different temperatures
for both the $k_3$ and $k_4$ parameters.
We clearly see that the lowest detectable anharmonicity 
is well below the measured one in both cases. In particular,
the rejection probability decays exponentially with increasing
magnitude of the anharmonic constant.

\section{\label{sec5} Conclusions}

In this paper we introduced a pair distribution function valid in the 
quantum regime based on a Taylor expansion of the absorber--neighbour potential,
suitable for the analysis of single--scattering EXAFS. 
In particular, we used ordinary  time--independent non--degenerate 
quantum perturbation theory to cast the QPDF in a simple analytical form, 
which can be easily calculated in a subroutine incorporated in the fitting
program. Moreover, we showed how the limits of validity (sensibility
of the quantum treatment and perturbative hypothesis) can be
monitored and we described how to implement a simple statistical test to
estimate the sensitivity of the QPDF to the third-- and fourth--order
terms in the potential. The latter procedure can be used {\em a posteriori}
to check whether the best--fit values of the anharmonic parameters 
are above the sensitivity level (minimum detectable anharmonicity).
Alternatively, the same procedure may be applied {\em a priori} 
in order to assess whether the QPDF is suitable for the analysis 
of the problem at hand.
We applied our formalism to the case of Silver Iodide, showing how
the potential anharmonicity can be measured in a temperature range
where the classical expression of the PDF could not be used.
In particular, we demonstrated how a simple harmonic model is not 
adequate to describing the dynamics of the Ag--I pair, in agreement
with previous studies performed in the classical regime.

As a final remark, it should be noted that the example we chose to test the
QPDF concerns a very symmetric lattice structure. It is well known that  an
additional broadening of the distributions of absorber--neighbour distances may
be produced by static disorder. The latter may be associated for example  with
the presence of a  coordination shell made of $N$ identical atoms at slightly
different distances from the photoabsorber, that can not be resolved in
different subshells (e.g. $N_1$ at distance $r_1$ and $N_2$ at distance $r_2$,
with $N_1+N_2=N$).  For small static disorder ($|r_1-r_2| \ll r_1,r_2$), one
can prove that such shell is equivalent to a shell with coordination $N$ and
mean distance $r_0 = (N_1r_1+N_2r_2)/N$, provided one introduces 
in the Debye--Waller factor the
additional  variance
$\sigma^2_{\rm stat}=N_1N_2|r_1-r_2|^2/N^2$~\cite{teo}.
In the framework of our model, this is expected to correspondingly 
rescale the harmonic constant $k_2$. However, it should not alter in general  
the information carried by the EXAFS signal regarding the anharmonic
terms of the absorber--neighbour effective potential.  
Hence, we expect that our model of lattice
anharmonicities may be sensibly used  in its present form also
in the presence of small static disorder.

Concluding, we developed an easy technique to study
interatomic potentials from single--scattering EXAFS in the quantum 
regime of atomic vibrations. A full--featured FORTRAN package
containing the relevant programs is made available by the author
to the interested scientists upon request. 

\section{Acknowledgments}
The author would like to thank Dr Paolo Ghigna for providing the 
EXAFS spectrum analysed in the present work and, more importantly,
for his constant help and support. The author is also indebted 
to Dr Luciano Cianchi for numerous precious discussions.

This work has been partly supported by Heriot--Watt University,
Edinburgh, during the Ph.D. course of the author under the
supervision of Dr Eitan Abraham.
\newpage
\section*{References}
\addcontentsline{toc}{section}{References}          


\newpage
\begin{table}
\begin{center}
\caption{\label{tab_alph}}
\begin{tabular}{c c c}
\hline\hline
$\s$  & $\a^{+}(n,\s)$ & $\a^{-}(n,\s)$ \\
\hline 
1 & $3(n+1)$  & $3n$      \\
2 & $2(2n+3)$ & $2(2n-1)$ \\
3 & 1         & 1         \\
4 & 1         & 1         \\
\hline\hline
\end{tabular}
\end{center}
\end{table}
\begin{table}
\centering
\caption{\small \label{tab_AgI_mia} 
Best--fit values of of the fit index~(\ref{resid}) and
parameters describing the Ag--I potential. The weight functions 
used are here $w_i=1/\sqrt{50}$ $\forall \, i$.}
\smallskip
\begin{tabular}{l | c | c c c c c c}
\hline \hline
model & ${\cal F}_{min}$ & $k_2$ (eV \AA$^{-2}$) & $k_3$ (eV \AA$^{-3}$) & $k_4$
(eV \AA$^{-4}$) &
$\eta$ (\AA$^{-2}$)& $r_0$ (\AA) & $\Delta E_0$ (eV)\\
\hline	
$\{ \l \}$        &  41.0 & 2.11(3) &   -     &  -    & 0.84(5) & 2.86(1) & -48(1)  \\ 
$\{ \l^{\pr} \}$  &  31.5 & 1.86(3) & -7.6(4) & 49(6) & 0.82(5) & 2.87(1) & -43(1)  \\ 
\hline \hline
\end{tabular}
\end{table} 
\begin{table}
\centering
\caption{\small \label{tab_AgI} 
Ag--I potential parameters as measured in ref.~\cite{Yoshi2}.}
\smallskip
\begin{tabular}{l r r r}
\hline \hline
$T$(K) & $k_2$ (eV \AA$^{-2}$) & $k_3$ (eV \AA$^{-3}$) & $k_4$ (eV \AA$^{-4}$) \\
\hline
300 & 2.4(1) & -5.0(1) & 0.2(0.7)  \\
600 & 2.7(1) & -5.3(1) & 0.08(0.4) \\
\hline \hline	     
\end{tabular}
\end{table} 

\end{document}